# Strain-Compensated AlInGaAs-GaAsP Superlattices for Highly-Polarized Electron Emission[*]

A. V. Subashiev, L. G. Gerchikov, Yu. A. Mamaev, Yu. P. Yashin

*Experimental Physics Department, State Polytechnic University,*

*Politekhnicheskaya 29, St.Petersburg 195251, Russia*

J. S. Roberts

*Department of Electronic Engineering, University, Sheffield S1 3JD, UK*

D.-A. Luh, T. Maruyama, and J. E. Clendenin

*Stanford Linear Accelerator Center, Menlo Park, CA 94025, USA*

## Abstract

Spin-polarized electron emission from the first superlattice photocathodes developed with strain compensation is investigated. An opposite strain in the quantum well and barrier layers is accomplished using an InAlGaAs/GaAsP superlattice structure. The measured values of maximum polarization and quantum yield for the structure with a 0.18 $\mu$m-thick working layer are close to the best results reported for any strained superlattice photocathode structure, demonstrating the high potential of strain compensation for future photocathode applications. An analysis of the photoemission spectra is used to estimate the parameters responsible for the polarization losses.

Submitted to Applied Physics Letters.

---

[*] This work supported in part by the U.S. Department of Energy under contract DE-AC02-76SF00515.

Strained short-period superlattice (SL) structures as candidates for the photocathodes of highly spin-polarized electron sources have been the subject of a number of studies; e.g., see refs. [1-3]. In these structures, the heavy-hole (hh) and light-hole (lh) minibands are split due to the difference in the hh and lh confinement energies in the SL quantum wells, which adds to the separation due to strain alone. The enlarged valence band splitting results in a high initial electron polarization in the conduction band under excitation by circularly polarized light. However, the thickness of the stressed photocathode working layer necessary for achieving a high value of quantum yield exceeds the critical thicknesses for strain relaxation, which results in structural defects, smaller residual strain and thus lowers polarization. A second factor limiting the maximum polarization is the smearing of the interband absorption edge, which is mainly due to the valence band tails and to the hole scattering processes.[4] Consequently the polarization in the bandedge absorption is less than 100 %, polarization losses being typically about 6%.[5]

To overcome these problems the use of strain compensation was proposed,[6] whereby the composition of the SL barrier layers is chosen to have opposite (tensile) strain from that of the quantum well layers. Due to the lowered barriers for the light holes formed by the tensile-strained barrier layers, the resulting valence band splitting in the strain compensated structures is expected to be smaller than in the structures without strained barriers but with similarly strained wells. However, considerably larger strain values in the SL wells can be achieved with no limitations on the overall thickness of the SL structure,[7] which should ensure high electronic polarization.

In this letter, we report the results of the first experimental and theoretical studies of polarized electron emission from newly-developed strain-compensated InAlGaAs-GaAsP SLs. The viability of strain compensation for polarized electron sources is demonstrated. In addition, a comparison of the experimental and calculated polarization and quantum yield spectra enabled us to determine the absorption-smearing parameters for the edges of the hole minibands and also indicated that the surface band bending region (BBR) makes a crucial contribution to the near-band edge absorption and to the maximum polarization value.

The design of the working layer SL was based on the band edge line up across the heterointerface between the compressively and tensile strained layers.[8] For strained SLs grown on GaAs substrates, the positions of the layer band edges can be calculated using the positions of the band edges in the GaAs layer as a reference. The range of available compositions of the $In_xAl_yGa_{1-x-y}$ As QW layer is restricted to $x$ values that give a maximum valence band splitting while retaining a high structural quality (i.e., $x \approx 16$-18%) and to $y$ values that result in a SL band gap larger than that in GaAs (i.e., $y \geq 12\%$).

The choice for the composition of the $GaAs_{1-z}P_z$ barrier layer is more complicated. For small $z$ values the tension in the barriers is not sufficient to compensate the deformation of the well layers, while the average valence band offset is too small to provide a large valence band splitting. For large $z$ and $y$ the enlarged valence band splitting in the barriers makes a type II SL for the light holes, which lowers the confinement energy of the light hole (lh1) miniband and the hh1-lh1 splitting.

The SL structures were grown by metal-organic vapor phase epitaxy (MOVPE), which presents several potential advantages, the main being high structural quality and low defect densities at the interfaces. In addition, the growth of phosphide materials is greatly simplified, and a wide range of growth rates and composition variations is achievable.[9] To separate absorption in the working layer from that in the substrate, a 0.5 $\mu$m-thick $Al_{0.3}Ga_{0.7}As$ $p$-doped buffer layer was grown on a $p$-type (100) GaAs substrate with a 3° mis-cut angle towards [110]. The working layer consisted of 8 to 30 pairs of compressively-strained $In_xAl_yGa_{1-x-y}As$ quantum-well layers alternating with tensile strained $GaAs_{1-z}P_z$ barrier layers. On top of the SL working layer, a 6-nm thick GaAs surface layer was deposited with Zn-doping concentration enlarged from $5\text{-}7 \times 10^{17}$ cm$^{-3}$ in the working layer to $1 \times 10^{19}$ cm$^{-3}$. Three sets of samples with $z=8$, 17 and $z \geq 19\%$ respectively with various working-layer thicknesses and other minor differences were grown.

For the average valence band offsets calculated as recommended in ref. [8], the resulting band offsets (i.e., the well depths) for the light holes are rather small due to the valence band splitting in the barriers, which results in a shift of the lh1 band over a range of ~15 meV as the SL layer thickness is varied. The samples were characterized by photoluminescence measurements and also by (004) x-ray diffraction scattering.

For the analysis of the experimental spectra, we have calculated the miniband, optical absorption, quantum yield, $Y(h\nu)$, and the electron polarization spectra, $P(h\nu)$, for these structures using the multiband Kane model,[4,10] including the conduction band $\Gamma_6$, the states of light and heavy holes of the valence band $\Gamma_8$ and also the states of the spin-orbit splitted $\Gamma_7$ band.

The excitation spectra of the polarized photoemission were measured at room temperature for different temperatures, $T_h$, used to clean the surface before activation. The structures showed the features that are typical for polarized electron emission from SL structures, including a high-polarization peak at the band-edge absorption and a second peak at higher energies with a well pronounced dip between them.

The experimental $P(h\nu)$ and $Y(h\nu)$ spectra (for several $T_h$ close to optimal) for the sample with $z=17\%$ (20 periods of 5 nm-thick wells and 4-nm barriers) that exhibited the largest polarization value, $P_{max} = 84\%$, at the polarization maximum with $Y(h\nu_{max})= 0.4\%$ are presented in Fig. 1. These parameters are close to the best results reported for any strained superlattice photocathode structures, indicating the high potential of strain compensation.

The comparison of the experimental and calculated spectra showed that the maximum polarization value at the top of the first polarization maximum is sensitive to the smearing of the absorption edge associated with two main factors. The first is the interband absorption smearing due to the band edge fluctuations, which can be evaluated by fitting the $Y(h\nu)$ dependence allowing non-homogeneous Gaussian broadening of the absorption spectra. Such a fitting results in a smearing (band tail) energy parameter, $\delta$, of $\leq 15$ meV. The second factor originates from the processes of hole scattering between the hh and lh states, which leads to a non-zero contribution of the lh1 miniband to the absorption near the edge and that populates the second spin state. We have calculated the characteristic energy associated with the broadening of the light-hole absorption spectra, $\gamma$, in terms of the appropriate transition rate. The hole scattering rate on the acceptor impurity centers, estimated also from the mobility data for similar samples, is found to dominate and can result in $\gamma \approx 10$ meV, while the evaluated contribution of the other competing mechanism, scattering on the optical phonons, is $\gamma \leq 4$ meV at room temperature. The calculated spectra for $\gamma = 7$ meV and $\delta = 11$ meV are depicted in Fig. 1 by the dashed line giving the maximum value of electron polarization, $P_{max}$, of ~92%.

Finally, a sizeable contribution to the total photoabsorption at the polarization maximum should come from the surface GaAs layer forming the BBR and having a smaller bandgap. In structures with less phosphorus in the barriers, the surface layer may even be tensile strained, resulting in excitation of electrons with predominantly opposite spin direction. This contribution manifest itself as a sharp decrease of the polarization below the absorption edge since the polarization of electrons excited in the surface layer does not exceed 50%.

The results of the calculation of the *P(hv)* and *Y(hv)* spectra allowing absorption in a BBR layer with a thickness of 10 nm, assuming it to be unstrained, are shown in Fig. 1 by the solid line. The good agreement with experimental data shows the importance of this contribution to the total polarization losses. An additional argument favoring this mechanism is the substantial spread of the results with $T_h$ (at least ~3%), which presumably affects only the surface layer and distribution of the acceptors near the surface. The relative contribution of the surface layer depends on the doping profile and can be reduced in structures having a thinner BBR region due to a heavier surface layer doping and in structures with a thicker working layer.

The calculated hh1-lh1 splitting, $\Delta\varepsilon_{hh1-lh1}$, varies with the AlInGaAs well thickness from 27 meV to 44 meV. The observed variation of $P_{max}$ with the valence band splitting for all samples for different $T_h$ and the calculated dependence for various values of the smearing parameters are presented in Fig. 2. It shows that unrealistically high values of $\gamma$ and $\delta$ are required to get the calculated curve to even roughly fit the experimental data, while the correspondence is quite good when BBR absorption is allowed. Though the calculated dependence almost saturates above $\Delta\varepsilon_{hh1-lh1}$=40 meV, larger values of the splitting are preferable, since they should correspond to smaller smearing. Note that the thickness of the working layer of the sample with the highest polarization (0.18 $\mu$m) was considerably larger than that (0.1–0.12 $\mu$m) for the formerly used and studied strained-layer samples and strained-well superlattices with similar polarization values,[1-3] which directly confirms the validity of the strain compensation idea.

To summarize, for the first time photocathode superlattice structures with strain compensation have been grown and studied as candidates for highly polarized electron emission. These photocathodes are based on InAlGaAs-GaAsP structures grown by MOVPE. Allowing for the electron spin relaxation and weakly-polarized optical absorption at the surface, the calculated polarization spectra are in a good agreement with the observed excitation spectra of polarized electron photoemission. The prevention of strain relaxation and the smaller relative contribution of the BBR region in comparison with strained-well structures can make these structures advantageous.

This work was supported by CRDF under grant RP1-2345-ST-02, by NATO under grant PST.CLG 979966, by Russian Ministry of industry, science and technology under contracts # 40.012.1.1.1152 and # 40.072.1.1.1175, by UK EPSRC support for the National Centre for III-V Technologies at the University of Sheffield under grant GR/R65534/01, and also supported in part by the U.S. Department of Energy under contract DE-AC02-76SF00515.

**Figure Captions**

FIG.1. Polarization and quantum yield spectra of the emitted photoelectrons for $In_{0.18}Al_{0.12}Ga_{0.68}As$-$GaAs_{0.83}P_{0.17}$ superlattice for various pre-activation temperatures (discrete symbols) and the calculated energy dependence of *P(hv)* and *Y(hv)* spectra with the valence band smearing and hole scattering (dashed line) and with the contribution of the BBR added (solid line).

FIG. 2. Electron polarization, $P_{max}$, vs. the calculated splitting of the hh1 and lh1 valence-level minibands: experimental results (discrete symbols); dashed and dash-dotted lines show calculated results with the valence band smearing and broadening; solid line: results with allowance for BBR absorption.

**Figure 1. Authors: J. E. Clendenin et al.**

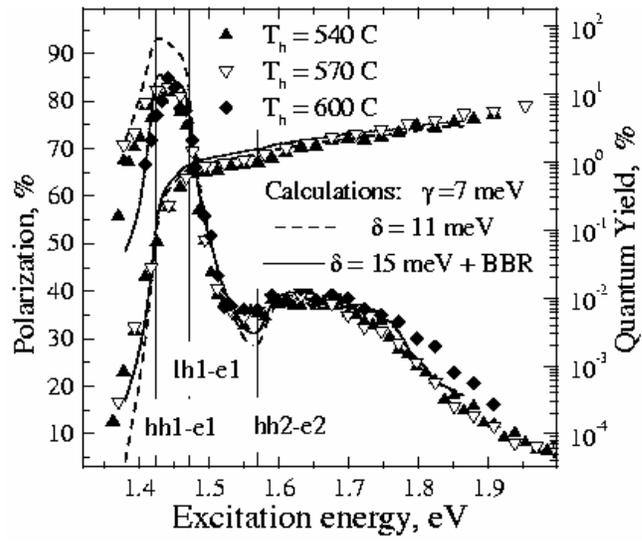

**Figure 2. Authors: J. E. Clendenin et al.**

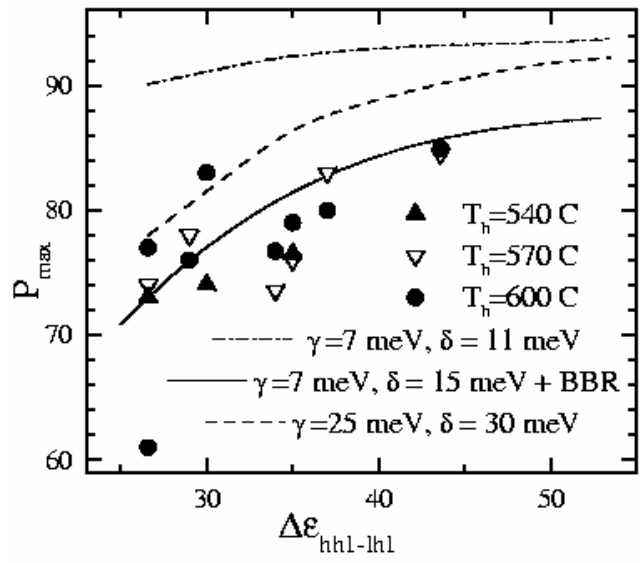